\documentclass [12pt]{MYIOPART}	% just a different name

\usepackage{epic}    
\usepackage{eepic}

\newcommand {\ba}      {\begin{array}}
\newcommand {\ea}      {\end{array}}

\newcommand {\be}      {\begin{equation}}
\newcommand {\ee}      {\end{equation}}

\newcommand {\bea}     {\begin{eqnarray}}
\newcommand {\eea}     {\end{eqnarray}}

\newcommand {\bes}     {\begin{displaymath}}
\newcommand {\ees}     {\end{displaymath}}

\newcommand {\beas}    {\begin{eqnarray*}}
\newcommand {\eeas}    {\end{eqnarray*}}

\newcommand {\ben}     {\begin{enumerate}}
\newcommand {\een}     {\end{enumerate}}

\newcommand {\bit}     {\begin{itemize}}
\newcommand {\eit}     {\end{itemize}}

\newcommand {\bt}      {\begin{table}}
\newcommand {\et}      {\end{table}}

\newcommand {\btb}     {\begin{tabular}}
\newcommand {\etb}     {\end{tabular}}

\newcommand {\bind}    {\begin{indented}}
\newcommand {\eind}    {\end{indented}}

\newcommand {\Ref}[1]  {(\ref{#1})}

\newcommand {\half}    {\textstyle {\frac {1}{2}}}

\newcommand {\fourth}  {\textstyle {\frac {1}{4}}}

\newcommand {\oonep}   {{\frac {1}{ \pi}}}
\newcommand {\otwop}   {{\frac {1}{2\pi}}}
\newcommand {\ofourp}  {{\frac {1}{4\pi}}}

\newcommand {\nid}     {\noindent}

\newcommand {\vk}      {\bi{k}}
\newcommand {\vK}      {\bi{K}}
\newcommand {\vn}      {\bi{n}}

\newcommand {\vY}      {\bi{Y}}
\newcommand {\vZ}      {\bi{Z}}

\newcommand {\vW}      {\bi{W}}
\newcommand {\vxi}     {\bi{\xi}}

\newcommand {\bg}      {\bar {g}}
\newcommand {\by}      {\bar   y}
\newcommand {\bz}      {\bar   z}

\newcommand {\bgxx}    {\bg_{_{X\,X\,}}} 
\newcommand {\bgxw}    {\bg_{_{X\,W  }}} 
\newcommand {\bgxz}    {\bg_{_{X\,Z\:}}} 
\newcommand {\bgww}    {\bg_{_{W  W  }}} 
\newcommand {\bgwz}    {\bg_{_{W  Z\:}}} 
\newcommand {\bgzz}    {\bg_{_{Z\:Z\:}}} 
\newcommand {\bgyy}    {\bg_{_{Y\:Y\:}}}

\newcommand {\CO}      {$C^{^0}\!$}
\newcommand {\CI}      {$C^{^1}\!$}

\newcommand {\CINF}    {$C^{^\infty}\!$}

\newcommand {\SII}     {$S_2$}

\newcommand {\HS}      {$\Sigma$}
\newcommand {\ST}      {$\cal S$}

\newcommand {\BC}      {$C_{\rho\sigma\tau}$}
\newcommand {\BCS}     {$C_{\rho\sigma'\tau'}$}

\newcommand {\sign}    {\mbox{signum}}

\newcommand {\vep}     {\varepsilon}

\newcommand {\corr}    {~\widehat{=}~}

\newcommand {\ie}      {i.e.~}
\newcommand {\eg}      {e.g.~}
\newcommand {\wrp}     {w.r.~}

\newcommand {\cmp}     {comp.~}
\newcommand {\rsp}     {resp.~}
\newcommand {\rhs}     {r.h.s.~}
\newcommand {\lhs}     {l.h.s.~}
\newcommand {\eal}     {et al~}

\newcommand {\mand}    {\quad \mbox{and} \quad}
\newcommand {\where}   {\quad \mbox{where} \quad}

\newcommand {\cart}    {\!\wedge\!}  % the symbol for outer (Cartan-) multiplication of forms
\newcommand {\cross}   {\!\otimes\!} % the symbol for tensor product of vectors
\newcommand {\cont}    {\!\cdot\!}   % the symbol for a single tensor-contraction
\newcommand {\dual}    {\!\star\!}   % the symbol for the Hodge-dual of a differential form
\newcommand {\tim}     {\!\times\!}  % the symbol for a cross-product

\newcommand {\Lap}     {\Delta}

\newcommand {\rem} [1] {} % to comment-out some text

\newcommand {\jump}[1] {\big[\, #1 \,\big]}
\newcommand {\avrg}[1] {\big(\, #1 \,\big)}

\newcommand {\cp}      {cut$\,\&\,$paste}
\newcommand {\CP}      {cut$\,\&\,$paste }
\newcommand {\RN}      {Reiss\-ner--Nord\-str\"om }
\newcommand {\ME}      {Max\-well--Ein\-stein }
\newcommand {\MESE}    {Max\-well--Ein\-stein sys\-tem of equa\-tions }
\newcommand {\Li}      {Lich\-ne\-ro\-wicz }
\newcommand {\li}      {Lich\-ne\-ro\-wicz}
\newcommand {\Mi}      {Min\-kow\-ski }
\newcommand {\Sc}      {Schwarz\-schild }
\newcommand {\DtH}     {Dray and 't Hooft }

\newcommand {\AaS}     {Aichel\-burg and Sexl }
\newcommand {\AS}      {Aichel\-burg--Sexl }

\newcommand {\MP}      {Ma\-jum\-dar--Pa\-pa\-pe\-trou }

\newcommand {\ray}     {Ray\-chaud\-hu\-ri}

\newcommand {\PV}      {Podolsk\'y and Vesel\'y }
\newcommand {\AB}      {Aichel\-burg and Bala\-sin }
\newcommand {\BN}      {Bala\-sin and Nach\-ba\-gau\-er }

\newcommand {\BH}      {Beken\-stein--Haw\-king }
\newcommand {\GaT}     {Ge\-roch and Trasch\-en }

\newcommand {\prtl}[1] {\partial_{_{#1}}}
\newcommand {\tbfn}[1] {\raisebox{1ex}{\,\footnotesize #1}}

\begin{document}

\begin{flushright}
UWThPh--2001--26 \\
July 2001
\end{flushright}

\title[Nonanalytic extreme \RN extensions in terms of weak solutions]
{
\begin{center}
Nonanalytic extensions \\
       of the extreme \RN metric \\
       in terms of weak solutions.
\end{center}
}

% \submitto{\CQG}

\author{Wolfgang Graf \footnote[7]{visiting scientist} }

\address{
Institut f\"ur Theoretische Physik, Universit\"at Wien, \\
Boltzmanngasse 5, A--1090 Wien, Austria }

\ead{graf-reichenau@t-online.de}

%  \date{\today}

\sloppy

\begin{abstract}

A basic extension of the exterior part of the extreme
\RN~solution in terms of a continuous metric and
gauge potential is constructed.
This extension is not smooth at the null hypersurface
given by the Cauchy--Killing horizon
which separates isometric copies of the exterior metric.
The \MESE~is satisfied only in a weak sense.
The manifold is topologically incomplete and the spherical symmetry 
is globally broken down to an axial symmetry.
This behaviour can be attributed to the effect
of a `topological string', in the sense of a infinitesimally thin closed
stringlike object `sitting on the rim' of the black hole
and holding it open by means of an accompanying
impulsive gra\-vi\-ta\-tio\-nal wave.
The resulting differentiable manifold and the corresponding 
horizons are not anymore simply connected,
being `pierced' by the strings.

\end{abstract}

\section{Introduction}

The standard extensions of the metrics describing the
stationary gravitational field of a massive, and possibly
electrically charged and rotating particle 
(with mass $m$, electric charge $p$ and specific
angular momentum $a=J/m$)
through its horizon  (when $m^2 \geq p^2 + a^2$)
are all based on an {\em analytic} continuation
of the metric in appropriate coordinates,
and leading to the family of metrics 
of the Kerr--Newman class,
parametrized by $(m,p,a)$.

As well known, these analytic extensions are riddled not only
with the peculiar phenomena related to the
existence of black boles, but more seriously,
with inevitable singularities in their interior,
where the mathematical description is put to stress.
Still, for the metrics under discussion, it was possible
to derive {\em distributional sources} at their 
singular locii (\cmp \eg \BN \cite{BaN93}, Balasin \cite{Bal97})
\smallskip

Both the Maxwell and the Einstein
equations, as well as their combination,
are hyperbolic (due to the Lorentzian
character of the space--time metric)
and so do admit nonanalytic, or wavelike solutions.
In fact, as any smooth lightlike hypersurface
could serve as a surface of discontinuity
(more technically called characteristic surface),
it would be a safe bet to claim that such
nonanalytic solutions are much more typical
than the purely analytic ones, in the sense
that they would make them appear to belong
to a set of zero measure in a presumed
space of solutions.

Of course, the significance of such non--smooth
solutions has not been completely ignored,
and there is a growing literature 
in the context of freely propagating
gravitational waves (\cmp \eg Griffiths'
monograph \cite{Gri91}).

However, it seems that up to now it has not
been attempted to make nonanalytic extensions
to the classical metrics mentioned above.
This is what we try do here for the
particular case of an extremal
\RN metric (\ie $m > 0,\,p = \pm m,\,a=0$).

In fact, the horizon \HS~of the extremal metric,
besides being doubly--degenerate, is also
a Killing horizon, in the sense that its
null--generator is a Killing vector $\xi$
for the metric restricted to it.
This situation allows some extra degrees
of freedom when cutting the metric along
its horizon and then pasting it again
with a copy of itself, after a relative shift
along $\xi$ of the corresponding points --- 
just a variant of Penrose's method 
of scissors--and--paste \cite{Pen72}\footnote[1]
{when in the following mentioning Penrose, 
we will always refer exclusively to this classical paper},
slightly generalized to cope with topologically
nontrivial situations.
Also we will be able to fill some minor
gaps (perhaps considered by him too obvious 
to need an explanation), 
by systematically exploiting some basic
properties of {\em local Killing horizons}.
Incidentally, such objects play a major role
in some recent attempts of Ashtekar et al
(\cmp \cite{Ash00}, \cite{ABL01}) to generalize
the notion of horizons, with the aim of
a better understanding
of the thermodynamic properties
of black holes and related objects.
\medskip

As the main application of the formalism developed,
we will show that it will be possible 
to reglue two copies of the external part of
the extremal \RN metric along \HS~in
such a manner, that the distributional part
of the Einstein tensor, supported by \HS,
vanishes.
For this the {\em reduced Einstein equation}
\mbox{$\Lap f - 2 = 0$} has to be satisfied
on a metric 2--sphere \SII.
To achieve a globally well--behaved extension seems 
to be at first out of question,
as this equation does not admit even a generalized solution
in terms of distributions.
\smallskip

However, we can readjust the global
topology, so that the basic solutions 
\mbox{$f=-2\,\ln(1\pm\cos(\vartheta))$}
of the reduced equation,
as well as the corresponding extension, 
still make sense.
This we can achieve by making a further
\CP surgery procedure, this time
along the equatorial plane
given by $\vartheta = 0$,
and performing independently certain coordinate
transformations, before regluing them back
without any shift. In this way,
no distributional curvature will appear at this join.
The metrics corresponding to these coordinate charts are
well--behaved only
on the corresponding hemispheres.
As a consequence, together with the
transitions through \HS,
there will now appear
four--cycles of transition functions,
which cannot anymore reduced to the
identity --- the locus
parametrized by $r=m$ (horizon)
and $\vartheta = 0$ (eq.~plane) has in fact
to be excluded from the resulting
smooth manifold.
Nonetheless, it can be enclosed by the
thin tube $(r/m - 1)^2 + \sin^2(\vartheta) = \vep^2$,
where the absolute value of its circumference
can be made as small as any positive number.
Therefore, we can speak of a {\em topological string}.
\medskip

Provided with this new differentiable structure,
it is the manifold itself, which must somehow be considered
as incomplete `at the string' --- in particular, any geodesic
`hitting the string', will also be incomplete.
Also, the mere presence of the string breaks down
the originally present spherical symmetry down
to an axial symmetry.
Although the distributional part of the Einstein
tensor vanishes, there will remain a nonvanishing
distributional part in the Weyl component of
the Riemann curvature tensor, supported by \HS.
Therefore our extension must be considered
an {\em impulsive gravitational wave}
in the sense of Penrose,
although it does not spread in space and time,
as it is bound to the stationary horizon.
This wave is carried along with the closed
string holding open the black bole.

Besides the metric, also the gauge potential
can be made continuous in a way that no
charged shells appear. However,
the Maxwell tensor will exhibit jumps.
Technically speaking, our construction, consisting
of the extended smooth manifold, a continuous
metric and continuous gauge potential,
can be considered as a {\em weak solution
in the sense of \li} of the \ME system
of equations.
A very brief introduction into 
weak solutions in the sense of \Li
will also be given.
\medskip

Some apparently closely related metrics
are the well--known metric of Aichelburg
and Sexl \cite{AiS71} \footnote[2]
{to be abbreviated with AS--metric} 
(obtained in the limit of a lightlike boost
of the \Sc metric and resulting in a thin sandwich wave),
and the metric of \DtH \cite{DrH85} \footnote[3]
{to be abbreviated with DH--metric} 
(with the interpretation of a null
particle sitting on the horizon of a \Sc metric),
which also describe particlelike structures
carrying along an impulsive wave. 
The most significant differences and similarities
with respect to our continuous extension
are briefly discussed. Although they might
be considered as weak solutions in some generalized sense,
they are definitely not so in the sense of \Li.
\medskip

In view that there is already some literature on
Strings in the context of General Relativity,
it seems appropriate to emphasize several aspects
which makes the present work not directly related to most of them:
\begin{itemize}
\item[i)] the `strings' considered here are only {\em stringlike
in a topological sense}, and in no way expressible by
strings of Nambu--Goto type as usually considered in String Theory.
\item[ii)] the `strings' considered here are obtained by the method of
\CP along the lines of Penrose \cite{Pen72}, and not by some
limiting procedure of a material source. In particular, 
{\em no `regularization procedure'} is involved, as would be the case
for the `classical strings' considered previously 
(\cmp \eg Geroch and Traschen \cite{GeT87}, 
Steinbauer \cite{Ste00}, Sj\"odin and Vickers \cite{SjV01}).
\item[iii)] the solution obtained is still an {\em exact solution}
of the full system of \ME equations, 
albeit in the Weak Sense of Distribution Theory.
In particular, no additional fields enter,
like dilaton fields.
\end{itemize}

\section{A `naive' Extension}

The line element of the extreme \RN metric \footnote[5]
{to be abbreviated with eRN--metric} 
in its standard form is 
\be
  \label{eRN}
  ds^2 = -H dt^2 + H^{-1} dr^2 + r^2 (d\vartheta^2 + \sin^2(\vartheta)\, d\varphi^2),
\ee
where $H = (1-m/r)^2$.
As corresponding Maxwell field two--form $F$ we take
\be
\label{Max}4\,\pi\,F = p/r^2\, dr\cart dt + q\,\sin(\vartheta)\,d\vartheta\cart d\varphi,
\ee
where the charges are restricted for extremality by $p^2+q^2=m^2$.

Evidently this form of the eRN--metric is regular 
(in the sense of being infinitely differentiable)
only for $r\neq0$ and $r \neq m$.
To overcome the corresponding incompleteness at the `\Sc radius' $r=m$
(where both the Maxwell--form and the curvature invariants are regular),
this metric must somehow be extended through it. 
The well--known standard extension due to 
Carter \cite{Car66} 
is based on the analytic extension of some particular 
metrics isometric with the original for $r>m$. 

By means of the coordinate transformations generated by the 
(for $r\neq m$) exact one--forms $du$ and $dv$, defined by 
\be
\label{Toutin}
m\, du = dt - H^{-1} dr \quad {\mathrm and} \quad m\, dv = dt + H^{-1} dr,
\ee
to outgoing \rsp ingoing
Eddington--Finkelstein coordinates, 
which themselves are related  for $r \neq m$ by
\be
\half\,m\,(dv - du) = H^{-1} dr,
\ee
the original eRN--metric
is extended in the past (\rsp future) null direction through its (degenerate)
Cauchy horizon at $r=m$, giving
\begin{equation}
\label{out}
m^{-2}\ ds^2_{(out)} = -\Big(\frac{x}{1+x}\Big)^2\ du^2 - 2\ du\, dx + (1+x)^2\ d\Omega^2,
\end{equation}
respectively
\begin{equation}
\label{in}
m^{-2}\ ds^2_{(in)} = -\Big(\frac{x}{1+x}\Big)^2\ dv^2 + 2\ dv\, dx + (1+x)^2\ d\Omega^2
\end{equation}
where the angular part $d\Omega^2$ of the metric is defined as
\begin{equation}
  d\Omega^2 =  \Theta\, dy^2 + \Theta^{-1}\, dz^2
\end{equation}
and we have set $x = (r/m-1)$, $\Theta = 1-z^2$,
$y = \varphi$ and $z = \cos(\vartheta)$.
As customary, we will tolerate the harmless breakdown of the angular coordinates
at $z=\pm 1$.
Also we pulled out the factor $m^2$, thus making the coordinates dimensionless.
Note that the exterior metric is given by $x \geq 0$ and also that the
corresponding horizons are located at $x=0$. 
Evidently these extensions are analytic through their horizons,
leading to the corresponding interior metric ($x \leq 0$).
By properly patching together the charts corresponding 
to such partial extensions, 
the complete analytic extension is then obtained.
Of course we could also have used Kruskal--like coordinates,
but their implicit character would make many of the steps
to be done relatively involved.
\medskip

Let us now glue together directly only the outer regions
of the two metric forms \Ref{out} and \Ref{in} at $x=0$, 
giving the particular form (or `na\"\i ve' extension)
\be
\label{metric}
m^{-2}\ ds^2 = -\Big(\frac{x}{1+|x|}\Big)^2\ dw^2 - 2\, \rho\ dw \, dx + (1+|x|)^2\ d\Omega^2,
\ee 
now also valid for $x\leq0$, and where $\rho=+1$ or $\rho=-1$, 
depending on the outgoing \rsp ingoing 
character of the metric for $x \geq 0$.
The side information of \HS~is given by
$\sigma := \sign(x)$,
so that $|x| = \sigma x$.
In this form the metric already consists of a continuous join 
at $x=0$ of the original metric
($\sigma = +1$) with the reflected copy ($\sigma = -1$)
for the whole $x$--range.

For $x \neq 0$ the relation between the ingoing and the outgoing coordinates
is now given by the differential relation between exact one--forms,
\begin{equation}
\label{Tinout}
  {\half}\,(dv - du) = \left(\frac{1+|x|}{x}\right)^2 dx,
\end{equation}
which is directly integrable to $v-u=2\,\sigma\,(|x|+2 \ln|x|-1/|x|)$

Formally\footnote[2] 
{in the sense of not bothering about the proper definition of an appropriate 
space of test functions}
using $(|x|)' = \sign(x)$ and $(\sign(x))' = 2\,\delta(x)$ 
(with the advantage of avoiding any of the much more 
involved glueing techniques),
the matter distribution of the above extended metric (for $\rho$ fixed),
results in the (covariant form of the) stress--energy tensor $e$ of the shell 
\HS~having the form of null dust with {\em negative} energy density,\footnote[3]
{a more detailed derivation is provided in the Appendix}  
\be
\label{dis}
e = -\oonep \, \delta(x) \, dx^2.
\ee
The negative energy density would be generally expected 
by invoking the Raychaudhuri--identity
for an initially contracting and then expanding family of 
radial null geodesics (being in addition affinely parametrized and 
vorticity--free) traversing the null hypersurface given by $x=0$.
Of course, in the classical context considered here, 
negative energy is considered to be highly undesirable,
and in the rest of this paper we will try to construct an alternative 
extension, where the null shell \HS~is devoid of {\em any} matter.

\section{Basics on Continuous Extensions}

Let us now look more closely at the hypersurface--metric 
\be
\label{h_metric}
d\sigma^2  := d\Omega^2
\ee
induced in the null hypersurface $\Sigma$ given locally by the 
smooth equation $x=0$.
Let us also stress, that \HS~is globally two--sided (\ie co--orientable).
Evidently, this \HS--metric is {\em degenerate}, being annihilated by 
the vector $\vk := \partial/\partial w$.
This characteristic vector $\vk$ is also (up to a factor) the restriction 
to the horizon of the non-spacelike Killing vector of the original metric
--- in fact, $\Sigma$ is a \emph{Killing horizon}.
\smallskip

The apparently innocent observation that with $\vk$, also the product 
$\lambda \vk$ is a 
characteristic Killing vector on \HS, 
will however turn out to be crucial for our extension,
as it allows an {\em isometric} remapping  
\be
w \to w + f
\ee
of the (degenerate) metric along the null generator $\vk$ of \HS,
where $f$ is {\em any} smooth function on \HS.
Such a null hypersurface \HS, intrinsically characterized 
by a degenerate metric and having a Lie--isometry
along its null generator $\vk$, is sometimes also called
{\em local Killing horizon} (\cmp Carlip \cite{Car99}).
In fact, the existence of a local Killing horizon 
is at the heart of Penrose's method of \cp.
This will be described more precisely in the following,
but adapted to our specific needs.
\medskip

To simplify later equations, 
the factors $\rho$ and $\sigma$ are deliberately introduced 
and $f$ is assumed not to depend on them, giving
\be
\label{pullback}
\gamma_f : w \to w + \rho\sigma\,f
\ee
Here the meaning of $\sigma$ is slightly extended
to also indicate the positive, \rsp negative side of \HS.
With this interpretation, eqn.~\Ref{pullback}
already describes succinctly the process of \cp:
the original `na\"\i ve' metric \Ref{metric} is cut along $x=0$ and the 
resulting two pieces reglued isometrically for a fixed $\rho$ with a 
side--dependent shift $\rho\sigma f$ \wrp to the affine parameter 
$w$ of the null generators $\vk$ of their respective bounding horizons.

As is well--known, an isometry between the \HS--metrics is necessary 
and sufficient for the continuity of the metric through \HS.
In fact, Clarke and Dray~\cite{ClD87} show
how to obtain in this case canonical coordinates
where the metric is manifestly continuous through \HS.
However, we will proceed more directly.
Assuming a `static gluing', generated by $f = f(y,z)$, 
a particularly simple set of transition functions,
manifestly preserving the continuity of the metric \Ref{metric}, is
\begin{equation}
\label{EF_trans}
{\boldmath {\Gamma_f : }}
\unboldmath
\left\{
\begin{array}{lcl}
  W &=& w + \rho\, \sigma\, (f + \half \,(\Theta^{-1} f_y{^2} + \Theta f_z{^2}) \ |x|) \\
  X &=& x \\
  Y &=& y + \Theta^{-1} f_y \ |x| \\
  Z &=& z + \Theta f_z \ |x|\,.
\end{array}
\right.
\end{equation}
Particular forms of this equation already appear in d'Eath \cite{Eat78}, 
Aichelburg and Balasin \cite{AiB96}, as well as in \PV \cite{PoV98}.
However, instead of our `symmetric' convention in terms of 
the absolute value function $|x|$, there
the `asymmetric' convention in terms of $x\,\Theta(x)$  is used,
where $\Theta$ is the Heaviside function.
\medskip

However it seems not to be known, that \eref{EF_trans} 
can be systematically derived.
Denote the (global) nonspacelike Killing vector by $\vk$.
Assume on the null hypersurface \HS,
given by the scalar equation $x=0$, that $\vk = a\,g^{-1}dx$,
with some appropriate scalar function $a$ positive on \HS. 
Then the vector
\be
\label{loc_K_hor}
\vK := \lambda\,\vk - a\,x\,g^{-1} d\lambda
\ee
which agrees with $\lambda\,\vk$ on \HS, 
is now also defined {\em away} from \HS~(a 
neighbourhood of \HS~would be sufficient).
It still satisfies on \HS~the Killing equation, 
but now \wrp to the {\em complete} metric.
For a given $\lambda$, there is still some freedom 
in the choice of $a$: any $\tilde a$,
which agrees with $a$ on \HS~would do.
With the choice $a=m^2$ and the assumption $\lambda(y,z)$, 
an equation like \Ref{EF_trans} 
can then be derived by exponentiation using $K$:
\be
\bi{X} = \exp(t \vK)\,\bi{x}
\ee
(where $\bi{X},\,\bi{x}$ denote the coordinate--columns),
after setting $t=1$ {\em and} discarding the $o(v^2)$--terms.\footnote[9]
{\ geometrically, $\boldmath {\Gamma_f}$ is a particular example 
of a `first order contact structure'}
Anyway, these terms would inherit the nonuniqueness from the factor $a$ ---
but more importantly, the results to be derived will depend 
only on the $o(v)$--part.
We will show in a moment explicitly how this mechanism works 
by way of a relatively simple example.
\medskip

Let us mention that the concept of a
{\em non--rotating isolated horizon}
introduced by Ashtekar \eal \cite{ABF00}, 
and recently generalized once more
to a {\em rotating isolated horizon} \cite{ABL01},
is closely related with the existence of a
{\em local Killing horizon}.
In fact, the existence of a LK horizon
(although not mentioned by name)
follows directly from their definitions
of NRI-- respectively RI horizons.
The infinite--dimensional algebra 
of symmetries for LK horizons has
been related to Virasoro--algebras
by Carlip \cite{Car99},
who on this basis derived
the \BH entropy for a variety of spacetimes.
Note that due to the somewhat stronger
definitions of NRI-- and RI horizons, 
the corresponding
symmetry group turns out to be one--dimensional,
{\em except} for extremal isolated horizons
(locally characterized by the vanishing
of their surface gravity).
Particular forms of the general expression 
for the Lie--generator \eref{loc_K_hor} 
of the local symmetry
are used as `Ans\"atze' in \cite{Car99} and in \cite{ABL01},
but without directly relating them to the basic
properties of a LK horizon.
\medskip

After this little disgression on other applications of LK horizons,
let us now continue with our version of Penrose's \cp.
The new coordinates $\{W,X,Y,Z\}$ make the resulting metric continuous
in a (sufficiently small) neighbourhood of $\Sigma$.
Evidently, $\Gamma_{\!f}$ goes over continuously to $\gamma_f$ for $x \to 0$.
The Jacobi determinant $J$ of the transformation $\Gamma_{\!f}$ 
is easily calculated, giving
\begin{equation}
  \label{jac}
  J_{\,\Gamma_{\!f}} = 1 + \Delta f\ |x| + Q(f,f) \ x^2,
\end{equation}
where $\Delta$ is the Laplacian on the 2-sphere with metric $d\Omega^2$,
\begin{equation}
  \label{lap}
  \Delta f :=  (\Theta^{-1} f_y)_y + (\Theta f_z)_z, 
\end{equation}
and the in $f$ quadratic operator $Q$ is defined by
\begin{equation}
  \label{quad}
  Q(f,f) := (\Theta^{-1} f_y)_y\,(\Theta f_z)_z  - (\Theta^{-1} f_y)_z\,(\Theta f_z)_y.
\end{equation}
For the reglued metric, which now explicitly depends on $|X|$,
either using the formal calculation as indicated above,
or applying the theory of null junctions 
(\cmp \eg Hamoui and Papapetrou~\cite{HaP68}),
the matter distribution $e$ (in its contravariant form) 
supported by \HS~can be shown to result in
\be
  e = \otwop \,(\Delta f - 2)\,\delta(x)\,\vk \cross \vk,
\ee
characterizing an impulsive null junction of Penrose type III.
Evidently, depending on the sign of $\Delta f - 2$, 
{\em locally} the distributional energy density can be given any sign
(note however, that it does {\em not} contribute to 
the Noether energy integral, 
as $\vk \cont e = 0$).
It can be even made to vanish if the following inhomogeneous Laplace
equation, the {\em reduced Einstein equation}, holds on the horizon \HS:
\be
\label{Lap}
\Delta f = 2.
\ee
Of course, the particular form of the reduced equation depends on
the two metrics being joined isometrically and on their join.
As just shown for the particular case of the join of two exterior \RN metrics,
it can however be derived systematically.
This question was left essentially open by Penrose.

\subsection{Example: general metric for a `continuous' pp--wave}

As an illustrative example, let us quickly derive the `continuous form'
of a general impulsive pp--wave obtained from joining two \Mi spaces
along a null plane.
In the following section we will discuss some of the problems
which prevent in general such extensions to be globally well--defined 
--- this is the reason that already here we use quotes 
for the term `continuous'.
\smallskip

Using pseudo--euclidean coordinates, the line--element
for \Mi spacetime can be written as
\bes
ds^2 = - 2\,du\,dv + \delta_{ij}\,dx^i\,dx^j, \where x^i = \{x,\,y\}.
\ees
Consider the null hypersurface \HS, given by $v=0$, 
with normal 1--form $k=dv$, thus resulting
in the normal vector $\vk = -\prtl{u}$. Evidently, 
the null vector $\vk$ is the restriction
of a global Killing vector to \HS. 
In particular, \HS~can be considered as a local Killing horizon.
Therefore the \CP method of Penrose is applicable here.
\smallskip

Define a generator of local (\ie \wrp to \HS) isometries by
\bes
\vK := -\lambda\,\prtl{u} - v\,\lambda_{i}\,\prtl{i},
\ees
and assume $\lambda=\lambda(x^i)$. Define $\lambda_i := \prtl{i}\lambda, \ \Lambda := \lambda^i\,\lambda_i$. 
Then 
\bes
\vK \left( \ba{c} u \\v \\x \\ y \ea \right) = -\left( \ba{c} \lambda \\ 0 \\ v\,\lambda_x \\ v\,\lambda_y \ea \right) 
\qquad \vK^2 \left( \ba{c} u \\v \\x \\ y \ea \right) = v\left( \ba{c} \Lambda \\ 0 \\ v\,\Lambda_x \\ v\,\Lambda_y \ea \right),
\qquad\mbox{etc.,}
\ees
giving as result of the exponential map
\bes
\left\{ \ba{rcl} 
U &=& u - \lambda\,t + \half \, vt^2\,\Lambda + o(v^2) \\
V &=& v \\
X &=& x - vt\,\lambda_x + o(v^2) \\
Y &=& y - vt\,\lambda_y + o(v^2) \,.
\ea  \right.
\ees
Ignoring the irrelevant terms $o(v^2)$ and making the map `side--dependent', 
as demanded by \cp, we then get for the `effective' map
\bes
\left\{ \ba{rcl} 
U &=& u + \sigma\,( \lambda\ + \half \, |v|\,\Lambda ) \\
V &=& v \\
X &=& x + |v|\,\lambda_x  \\
Y &=& y + |v|\,\lambda_y \,,
\ea  \right.
\ees
where $\sigma := \mbox{signum}(v)$. 
Pulling back the metric by it for each side separately
(thus ignoring any distributional derivative of $\sigma$)
and defining $\lambda_{ij} := \prtl{j}\prtl{i}\lambda$, we finally get
\bes
\label{gAS_cont}
ds^2 = - 2\,dU\,dV + (\delta_{ij} + |V|\,\Lambda_{ij})\,dX^i\,dX^j, \quad
\Lambda_{ij} := 2\,\lambda_{ij} + |V|\,\lambda^r_i\,\lambda_{rj} .
\ees
Up to minor notational differences, this metric is identical 
with the one given by \AB \cite{AiB96}.
Note however, that the derivation just given here differs essentially 
from theirs: 
whereas they derive it from a manifestly distributional metric 
(generalizing the \AS metric \cite{AiS71}) 
by a discontinuous coordinate transformation
and {\em take care} of the distributional derivatives 
of the discontinuities involved,
we just applied Penrose's method of \CP to \Mi space:
by transforming each half--space separately, and then reglue them again,
the derivatives of the discontinuities do not enter here.\footnote[3]
{although they manifest themselves when calculating the Riemann curvature}

Note, that as terms like the ill--defined squares of $V \delta(V)$ 
are ignored, their derivation 
cannot be considered as completely rigorous (this applies also to 
a similar derivation of Penrose) in the sense of classical distributions.
As shown by Steinbauer \cite{Ste00},
this deficiency can however be repaired in the context of 
Colombeau's theory of generalized functions.\footnote[1]
{just called `Colombeau distributions' in the following}

Now, the jump tensor, and the distributional parts of the 
Riemann-- and Ricci tensor straightforwardly result in
\bes
h_{ij} = 2\,\lambda_{ij}, \quad r_{ij\,kl} = k_{[i}\,h_{j][k}\,k_{l]}\,\delta(V), \quad 
r_{ij} = \Lap \lambda \: k_i k_j\,\delta(V) .
\ees
Evidently, for $\lambda \neq \mbox{const}$, there is an impulsive 
gravitational shock wave of Penrose type III.
In particular, where $\Lap \lambda = 0$, the shell \HS~is devoid of any matter.
Let us also give the expression for the Jacobi determinant, 
which will be  useful when calculating the volume density:
\bes
J := \det \, (\delta_{ij} + |V|\,\Lambda_{ij}) 
  = 1 + 2\,|V|\,\Lap \lambda + V^2 \Big( \lambda^{ij}\,\lambda_{ij} + (\Lambda_{xx}\Lambda_{yy} - \Lambda_{xy}^2) \Big).
\ees

\subsection{Generic Problems with `continuous' Extensions}

Of course, for a smooth function $\lambda(x,y)$, 
the resulting metric is also smooth,
except on \HS, where it is continuous.
However, we get an immaterial sheet supported by \HS, 
only when the reduced Einstein equation $\Lap \lambda = 0$ holds.
There are no nontrivial smooth solutions vanishing at infinity --- 
in fact, any (local) solution
vanishing at infinity must somewhere diverge. Strictly speaking, 
a nontrivial solution (vanishing at infinity)
always leads to a unbounded metric and so cannot really be 
considered as continuous.\footnote[2]
{for the behaviour of distributional solutions `at infinity', 
see \AB \cite{AiB98} and \cite{AiB00}}
However, there exist distributional solutions to the 
{\em distributionally generalized}
equation $\Lap \lambda = 4\pi \,p \, \delta_{E_P}$,
where the delta distribution is supported by a point $P$
of the $(x,y)$--plane $E$,
which we can take to be the origin $(x=0,y=0)$.
A generalized solution is given by  
$\lambda = \half \,p\,\ln(\frac{\rho}{\rho_0})^2$,
where $\rho^2 := x^2 + y^2$ and $p,\,\rho_0$ positive constants ---
the same function as for the manifestly distributional \AS metric \cite{AiS71}.
But this does not automatically mean that the corresponding 
metric \Ref{gAS_cont}
can also be considered as a distribution --- quite to the contrary.
As $\lambda$ is locally--integrable, it can be considered as a distribution
supported by a two--plane $E$. Then all the derivatives occurring 
in the expression
for the reglued metric also exist as distributions. 
However the term $\lambda^r_i\,\lambda_{rj}$ in $\Lambda_{ij}$, 
being a product of distributions,  
does not anymore make sense as a distribution (in the `classical' 
sense of Schwartz)
and so this metric {\em cannot} properly be understood as such.
Therefore this `continuous' form of the metric is neither strictly
continuous, nor does it make sense as a `classical' distribution.

Another generic problem for distributional sources 
supported by lower--dimensional manifolds of codimension $m > 1$
(\eg line sources) can be seen in the result of \GaT \cite{GeT87},
questioning their well--definedness as (classical) distributional
sources of the Einstein equations.\footnote[1]
{\cmp Steinbauer \cite{Ste00} for a detailed discussion}
\smallskip

Apart from the singular behaviour at $\rho=0$,
there is still another serious problem.
The null hypersurface \HS, the `null particle' on it
at $v=0,\,\rho=0$ and the $z$--axis at $\rho=0$ are all {\em shielded} 
from the rest 
of the (flat) manifold by a {\em volume--singularity}.
In fact, $\mbox{vol} := J^\frac{1}{2} = |1 - p^2\,V^2\,(\frac{\rho_0}{\rho})^4|$ 
{\em vanishes} on the two--sheeted hypersurface given by $\rho^2 =  |p\,V|\,\rho_0^2$,
and {\em diverges} on the $z$--axis for the points not on \HS.
It is not clear how to find a complete and consistent system of charts 
containing \HS~and 
transition functions, without any degeneracy of the volume element, 
but still preserving
the essential `continuity' properties of the metric ---
if this can be done at all.
\medskip

From this discussion it should be clear, that the apparently 
`innocent--looking'
continuous partner of the \AS metric is after all not so well--defined,
whereas the original \AS metric is well--defined as a (tensorial) distribution.
The original manifestly distributional form of the \AS metric should 
therefore be preferred
over the `continuous' one, unless the above mentioned problems 
are solved satisfactorily.
\medskip

It seems that the problems found might be generic for solutions of the 
Einstein equations obtained by the \CP method.
However, there are some exceptions to this expectation.
A notable exception is the choice $\lambda = x^2-y^2$
for the general pp--metric \Ref{gAS_cont}, leading to a continuous
join at \HS~and having an innocuous volume degeneracy well--separated from \HS.
This choice was made in one
of the explicit examples given by Penrose.
In view of these problems, 
each particular such metric `continuously extended' metric 
should be carefully checked for well--definedness.
This objection also applies to the results 
of \PV \cite{PoV98}, who on the contrary claim to have found 
an explicitly continuous form 
(essentially equivalent to the one of \AB \cite{AiB96})
for the metric of impulsive pp--waves.
\smallskip

So let us take up again our main thread,
where we want to show that the extreme 
\RN metric admits a continuous extension.

\section{Extensions with Null Particles  on the Horizon?}

For a differentiable function $f$, 
the reduced Einstein equation \Ref{Lap} for our eRN/eRN--join
cannot be solved globally 
on any of the the \SII~foliating the horizon \HS.
Nor does it admit a generalized solution on the whole \SII~in 
terms of a distributional $f$, unless the equation
is modified with a distributional `source' \eg as
\be
\label{D_Lap}
\Delta f - 2 = 4\pi\,\Big(\alpha\,\delta(1-z) + \beta\,\delta(1+z)\Big),
\ee
with constants $\alpha,\,\beta$ such that $\alpha+\beta = -2$.
The (up to an arbitrary additive constant) 
unique generalized solution (in the sense of being 
a distribution equivalent to a locally--integrable function over \SII)
would then be
\be
f = \alpha\ln(1+z)+\beta\ln(1-z),
\ee
formally resulting now in a nonvanishing Einstein--distribution
\be
  e = 4\pi\,\Big( \alpha\,\delta(1-z) + \beta\,\delta(1+z) \Big)\,\delta(x)\,\vk \cross \vk.
\ee
This is very similar to the distributional energy for the AS-- 
and for the DH--metric\footnote[1]
{the problem with $\delta^2$--terms in the original derivation
has in the meantime been solved by Balasin \cite{Bal00}},
and so one could be tempted to interpret the metric as being generated by 
two null {\em particles}
sitting at the poles of the $S_2$ generating \HS, with relative mass parameters
proportional to $\alpha$ and $\beta$. However, differently from the AS--limit
of the Schwarzschild--metric, and from the DH--metric, its contribution to the
oberver--dependent energy integral would always be {\em non--positive}
(but without contribution to the Noether energy integral).
More seriously, the metric is not anymore continuous in the strict sense, 
as it diverges at the locii of the null particles.
Moreover, it cannot even be understood as a classical tensorial distribution,
and the volume--density would diverge on the whole $z$--axis, 
except for $v = 0$,
where it is undefined.
As this form of extension exhibits all the potential problems
for `continuous' extensions discussed in the previous section,
we will therefore dismiss it.

\section{A New Differentiable Structure --  the `Topological String'}

Let us in the following consider the two independent solutions 
to the reduced Einstein equation \Ref{Lap},
\be
\label{sol}
f_\tau = -2\,\ln(1+\tau z), 
\ee
where $\tau=\pm 1$ distinguishes the upper cap $\hat S$ ($\tau=+1$)
from the lower cap $\check S$ ($\tau=-1$).
Evidently, they are smooth  for $z \neq -\tau$.
The corresponding nontrivial transition functions from the original 
EF--coordinates 
(labelled by  $\rho$) are then
\bea
\label{EF_to_X}
W^{\rho\sigma}_\tau &=& w - 2\,\rho\sigma \left(\ln(1+\tau z) - \frac{1-\tau z}{1 + \tau z}\, |x| \right) \\
Z^{\rho\sigma}_\tau &=& (1+2\,|x|)\,z - 2\,\tau\,|x|.
\eea
Note, that as they depend only on $z$,
the quadratic term $Q$ in the Jacobi determinant is trivially zero 
and we are left with
\be
J^{\rho\sigma}_\tau = 1 + 2\,|x|,
\ee
which is smooth and positive
both for $\sigma=+1$ and $\sigma=-1$ separately.
In particular, the corresponding Jacobian matrix will be always invertible
(in its domain of definition) --- thus leading to corresponding
well--behaved transition functions for $X \neq 0$.

Considering the eight different types of charts \BC, 
labelled by $(\rho,\sigma,\tau)$
and derived from the original EF--charts (labelled by $\rho$) by means of the
transition functions \Ref{EF_trans} based on the solutions \Ref{sol},
their corresponding transition functions are now {\em defined} as follows:
\smallskip

\nid
a) `radial' transitions $\boldmath{S}$ through the horizon \HS~($X=0$),
corresponding to a change only of $\sigma$:
\be
\label{rad_trans}
{\boldmath {S^\rho_\tau : }}
\unboldmath
\left\{
\begin{array}{lcl}
W^{\rho-\sigma}_\tau &=& W^{\rho\sigma}_\tau \\
X^{\rho-\sigma}_\tau &=& X^{\rho\sigma}_\tau \\
Y^{\rho-\sigma}_\tau &=& Y^{\rho\sigma}_\tau \\
Z^{\rho-\sigma}_\tau &=& Z^{\rho\sigma}_\tau \ .
\end{array}
\right.
\ee
\smallskip

\nid
b) `equatorial' transitions $\boldmath T$ through the equatorial plane ($z=0$)
corresponding to a change only of $\tau$
(note that in their corresponding overlaps, always $z \neq \pm 1$):
\be
\label{eq_trans}
{\boldmath {T^{\rho\sigma} : }}
\unboldmath
\left\{
\begin{array}{lcl}
W^{\rho\sigma}_{-\tau} &=& W^{\rho\sigma}_\tau +  2\rho\sigma\tau F(|X|,z) \\
X^{\rho\sigma}_{-\tau} &=& X^{\rho\sigma}_\tau \\ 
Y^{\rho\sigma}_{-\tau} &=& Y^{\rho\sigma}_\tau \\
Z^{\rho\sigma}_{-\tau} &=& Z^{\rho\sigma}_\tau + 4\tau\,|X|\,,
\end{array}
\right.
\ee
where 
\bes
F(|X|,z) := \ln\frac{1 - z}{1 + z} + 4\, \frac{z}{1-z^2}\, |X| \mand
z := {\displaystyle \frac{Z^{\rho\sigma}_\tau + 2\tau\,|X|}{1+2\,|X|}} \,.
\ees
\smallskip

\nid
c) for completeness, we should also mention the `in/out' transitions 
$\boldmath R$, 
obtained from \eref{Tinout}, where only $\rho$ changes
(of course, $X \neq 0$ in their overlap)
\be
\label{ino_trans}
{\boldmath {R^\sigma_\tau : }}
\unboldmath
\left\{
\begin{array}{lcl}
W^{-\rho\sigma}_\tau &=& W^{\rho\sigma}_\tau + 2 \sigma \left( |X| + 2\,\ln|X| - |X|^{-1} \right)\\
X^{-\rho\sigma}_\tau &=& X^{\rho\sigma}_\tau \\
Y^{-\rho\sigma}_\tau &=& Y^{\rho\sigma}_\tau \\
Z^{-\rho\sigma}_\tau &=& Z^{\rho\sigma}_\tau \ .
\end{array}
\right.
\ee
These transitions will however only be needed when we later discuss
the global extension.

Let us now consider only a system of charts covering 
{\em one particular horizon} \HS,
characterized by a fixed $\rho$.
Up to now, the domains of definition of the charts \BC~have 
not yet been defined ---
as images of the original EF--charts, 
in principle they could completely overlap, 
except on the poles $z=\pm1$.

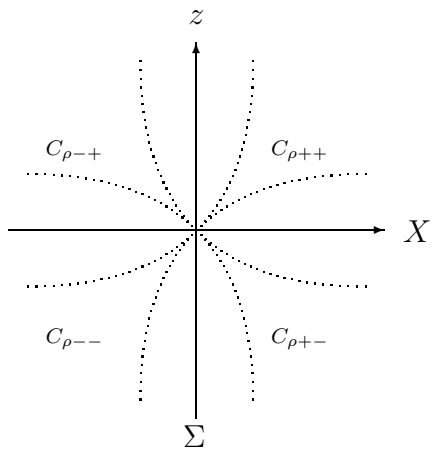
\begin{figure}[h]

\setlength{\unitlength}{0.5cm}

\begin{picture}(12,12)(-5,-1)

\thinlines
\put( 0,5){\vector( 1, 0){10}}
\put( 5,0){\vector( 0, 1){10}}

\qbezier[50](0.5,3.5)(6.5,3.5)(6.5,9.5)
\qbezier[50](3.5,0.5)(3.5,6.5)(9.5,6.5)
\qbezier[50](0.5,6.5)(6.5,6.5)(6.5,0.5)
\qbezier[50](3.5,9.5)(3.5,3.5)(9.5,3.5)

\put(1,7){$\scriptstyle C_{\rho-+}$}
\put(1,2){$\scriptstyle C_{\rho--}$}
\put(7,7){$\scriptstyle C_{\rho++}$}
\put(7,2){$\scriptstyle C_{\rho+-}$}

\put(4.65,-0.75){\HS}

\put(4.8, 10.5){$z$}
\put(10.5, 4.7){$X$}

\end{picture}

\caption{\label{horcharts} charts covering a horizon \HS}

\end{figure}

Now comes the {\em crucial point:} in order that all combinations of transition
functions are consistent in the overlap \BC~$\cap$ \BCS~of their charts, 
some point sets 
have to be {\em excluded} from them.
For example, assuming a fourfold overlap (for a fixed $\rho$), 
we would get the {\em nontrivial} four--cycle of transitions
\bes
W^{\rho\sigma}_\tau \stackrel{S^{-1}}{\to} W^{\rho-\sigma}_\tau \stackrel{T^{-1}}{\to} W^{\rho-\sigma}_{-\tau} 
\stackrel{S}{\to} W^{\rho\sigma}_{-\tau} \stackrel{T}{\to} \widetilde W^{\rho\sigma}_\tau 
= W^{\rho\sigma}_\tau + 4\rho\, F(|X|,z) \not \equiv W^{\rho\sigma}_\tau \! .
\ees
Such a situation can be avoided by excluding some points from 
the resulting manifold --- 
of course, we want to exclude as few points as possible.
A possible choice is to exclude the images of the 
two--dimensional point set \ST,
defined by \mbox{$x=0,\,z=0$}, from all the charts \BC.
This can be achieved \eg by the restrictions
\bes
(\sigma X + 1)(\tau z + 1) > 1, \where (\sigma X + 1) > 0, \quad (\tau z + 1) > 0,
\ees
now fixing their domains of definition.
With this choice, for a fixed $\rho$ any chart \BC~overlaps with exactly 
two others without having common triple points, 
thus avoiding any incompatibility of their transition functions.
This system of charts and their overlaps is sketched in \fref{horcharts}.

For $-\rho$ we get an independent system of charts and transition functions
for the other kind of horizon.
Evidently, this particular choice is not unique:
singling out on \HS~any other $z \neq 0$ with $\pm 1$ would also work.
More generally, any continuous and nonintersecting curve on 
\SII~separating its poles would still work, 
when Lie--transporting it along \HS~with $\vk$.
\medskip

Having to exclude a point set \ST~from the diffe\-ren\-tiable structure,
the differentiable manifold must be considered as incomplete,
in the sense of being topologically nontrivial there.
In all other respects the transition functions are well--behaved
and together with their charts can even be completed to an \CINF~atlas --- 
so we still have a {\em smooth manifold}, 
but provided with a {\em continuous metric}.
This spacetime must be considered {\em geodesically incomplete}, 
as there are geodesics which cannot be smoothly continued.
However the obstruction must be considered as purely {\em topological} 
as it still holds with respect to the \CO~level of the transition functions
characterizing a topological manifold.
\medskip

This obstruction to completeness can be enclosed by a hypersurface
locally given by the equation $x^2 + z^2 = \vep^2$
(where the positive constant $|\vep|$ can be made
arbitrarily small), with the topology $S_1 \tim S_1 \tim R$.
In the limit $\vep \to 0$ we then get a point set
with the topology of a {\em closed string}, $S_1 \tim R$,
where the factor $R$ corresponds to the path of the null vector $\vk$.
Even if this stringlike structure does {\em not} 
properly belong to the manifold,
let us in the rest of this paper simply call it {\em string}
and denote it with \ST.
\medskip

Evidently, the null hypersurface \HS, which originally
can be considered to be a product $S_2 \tim R$,
is now split by \ST~into $(H^+_2 \tim R) \cup (H^-_2 \tim R)$,
where $H^\pm_2$ are the open hemispheres corresponding
to a \SII~split by its equator.

Also note, that for each side $\sigma$ separately
the manifold can be made a smooth manifold with boundary \HS~--- 
even including the string $\cal S$.
However, this cannot be done simultaneously also 
on the other side $-\sigma$.
\medskip

We can also show that there is a direct {\em gravitational effect}
produced by~\ST, by calculating its {\em infinitesimal holonomy} 
(in the sense of parallel transport around a small loop)
around it as follows. For this we use a trick, which allows us
not to use the affine connexion. Like in flat space with a particular
symmetry axis $A$, instead of actively going around $A$,
we can essentially do it by a passive coordinate transformation.
The justification lies in the fact that we assume the limit
$\vep \to 0$ to be uniform, thus allowing to do this limit first.
Then we apply the nontrivial four-cycle of transformations
mentioned above, finally getting for the holonomy matrix
of an $n$--fold infinitesimal loop around \ST,
\bes
\mathbf T^n = \mathbf 1 
+ \lambda\, (\Theta^{\frac{1}{2}} dX \prtl{Z} + \rho\, \Theta^{-\frac{1}{2}} dZ \prtl{W}\!) 
+ \half\,\lambda^2 \rho\,dX \prtl{W} \where \lambda := 8\,n,
\ees
showing that the holonomy group $H$ for such loops is isomorphic to $\mathbf Z$.
This is {\em different} from the infinitesimal holonomy group for the cone 
with deficit angle on the $z$--axis,
where it is either periodic (rational angle) or nearly--periodic (otherwise).

By a similar analysis it can be shown, that a radial null geodesic $\prtl{X}$ 
infalling through \HS~near the equatorial plane
picks up a component $4\,\sign(z)\,\Theta^{\frac{1}{2}}\, \prtl{Z}$,
making the string \ST~gravitationally {\em repulsive}.
Again invoking \ray, this behaviour would normally be taken 
as a ma\-ni\-festation of negative energy. 
However with our extension,
this otherwise undesirable negative energy is so to speak
`swept under the carpet' of the string \ST,
which in fact is not even a proper part of the manifold ---
there is a deflecting effect without corresponding
matter being present as a localized source.
Extending a catchword of Wheeler coined for Geometrodynamics 
\cite{Whe62},
we can describe this situation succinctly as {\em matter without matter}.
\medskip

Note that this stringlike object differs from the  
commonly used notion of string in the original sense 
of Nambu--Goto, as it cannot properly be assigned
a two--dimensional surface (therefore the notion 
of `action' would be problematic). 
For a recent introduction into `standard'
gravitating strings, see Jassal and Mukherjee \cite{JaM01}.
However they use the term `null string' differently
to denote a NG--string with vanishing string tension.
\medskip

It is perhaps instructive to point out some of the main
differences and similarities of the metrics of \AaS 
and of \DtH 
with our continuously extended extremal \RN metric,
which we summarize in \tref{maindiff}.

\bt[h]
\caption{\label{maindiff} Main differences of the AS/DH--metrics vs.~the ceeRN--metric.}

\btb{@{}cllllll} \br
metric\tbfn{a}  & NP\tbfn{b}  & source\tbfn{c} & topology\tbfn{d} & inf. hol.\tbfn{e} & deflection\tbfn{f} & complete\tbfn{g} \\ \mr
AS/DH           & 1d op.      & yes            & trivial          & divergent         & attr./rep.         & incompl.  \\
ceeRN           & 2d cl.      & no             & nontrivial       & bounded           & repulsive          & incompl.  \\ 
\br
\etb
\smallskip

\nid
%%% maybe there is a better way to place some footnotes
{\footnotesize 
\tbfn{a} kind of metric: AS--, DH-- or continuously extended eRN--metric \\ [-1ex]
\tbfn{b} null particle is one--dimensional open vs.~two-dimensional closed \\ [-1ex]
\tbfn{c} localizability of the source of the corresponding NP \\ [-1ex]
\tbfn{d} topology in the vicinity of the NP \\ [-1ex]
\tbfn{e} infinitesimal holonomy of the corresponding NP \\ [-1ex]
\tbfn{f} whereas the NP for the ceeRN--metric is always repulsive, 
         for both the DH-- \\ [-1ex]
\tbfn{ }  and the AS--metric it can be made either 
         attractive or repulsive (free parameter) \\ [-1ex]
\tbfn{g} geodesic completeness \wrp to the NP 
}
\et

Being quite different in most aspects, the metrics discussed agree only
in being geodesically incomplete (for different reasons) at the null particle.
Whereas the DH--metric still has in addition the singularity at $r=0$,
such a singularity is not present for the ceeRN--metric.
The AS--metric, being flat except for the locus of the impulsive wave,
does of course also not have such a singularity.

\section{The Continuous Metric}
\label{ceern_metric}

Let us now write down the metric in a chart \BC~characterized by 
the set $(\rho,\sigma,\tau)$.
Pulling out the common factor $m^2$ from the metric: $g = m^2 \, \bg$, 
\beas
\bgxx &=& -4\,\frac{\Lambda\,|X|}{(1+2\,|X|)^2} \left( 2\,(1+2\,|X|)\,\Lambda - |X|
+ \left(\frac{1+2\Lambda\,|X|}{1+|X|} \right)^2 \Lambda\,|X| \right) \\
\bgxw &=& -\rho \left( 1 + 2\,\frac{1+2\,\Lambda\,|X|}{1+2\,|X|}\frac{\Lambda\,X^2}{(1+|X|)^2} \right) \\
\bgxz &=& -2\,\sigma\tau \, \frac{|X|}{(1+\tau z)^2} \left( \frac{2 - \left( (1+\tau z)-4 \right)\,|X| }{(1+2\,|X|)^2} \right. \\
 && \qquad \qquad \qquad \qquad {} + \left. 2\,\frac{(1+2\,\Lambda\,|X|)((1+\tau z)+2\,|X|)}{(1+|X|)^2(1+2\,|X|)^2}\,\Lambda\,|X| \right) \\
\bgww &=& -\frac{X^2}{(1+|X|)^2} \\
\bgwz &=& -2\,\rho\sigma\tau \, \frac{(1+\tau z)+2\,|X|}{(1+\tau z)^2(1+|X|)^2(1+2\,|X|)} \, X^2\\
\bgzz &=& \frac{1}{1-z^2}\left(\frac{1+|X|}{1+2\,|X|}\right)^2 
- 4\left(\frac{(1+\tau z)+2\,|X|}{(1+\tau z)^2(1+|X|)(1+2\,|X|)}\right)^2 X^2 \\
\bgyy &=& (1-z^2)(1+|X|)^2.
\eeas
Here we have used
\bes
\Lambda := \frac{1-\tau z}{1+\tau z} \mand z := \frac{Z+2\,\tau\,|X|}{1+2\,|X|}.
\ees
As immediately evident, the leading terms, including $o(|X|)$, are
\beas
\bgxx &\approx& -8\,\Lambda^2\,|X| \\
\bgxw &\approx& -\rho \\
\bgxz &\approx& -4\,\sigma\tau\,|X|(1+\tau z)^{-2} \\
\bgww &\approx& 0 \\
\bgwz &\approx& 0 \\
\bgzz &\approx& (1-2\,|X|)(1-z^2)^{-1}\\
\bgyy &\approx& (1+2\,|X|)(1-z^2).
\eeas
From this we obtain as the only nonvanishing jump expressions
\beas
\label{metric_jumps}
\jump{  \bgxx} &=& -8\,\Lambda^2 \\
\jump{  \bgxz} &=& -4\,\sigma\tau\,(1+\tau z)^{-2} \\
\jump{\,\bgzz} &=& -2\,(1-z^2)^{-1}\\
\jump{\,\bgyy} &=& +2\,(1-z^2).
\eeas
Defining the jump tensor $h$ 
as $h := \jump{g} = m^2\,\jump{\bg}$,
the distributional part of the Riemann curvature tensor
can be written as (\cmp \eg Hamoui and Papapetrou \cite{HaP68})
\be
  r_{ij\,kl} = n_{[i}\,h_{j][k}\,n_{l]}\,\delta(x), \where n_i \corr dx.
\ee
Evidently, the $\jump{\bgxx}$-- and $\jump{\bgxz}$--components drop out by the 
anti\-symmetrization involved with $r$
(in fact, any $dX$--component would drop out).
As an extra consistency--check, note that as required, 
the distributional part of the Ricci tensor vanishes:
$r_{ik} := g^{jl}\,r_{ij\,kl}=0$ ---
of course, as a consequence the distributional Einstein tensor 
\mbox{$e_{ij} := r_{ij} - \half r\,g_{ij}$} also vanishes.

Using the normalized 1--forms \mbox{$\by := m\,\Theta^\frac{1}{2}\,dy$}, 
\mbox{$\bz := m\,\Theta^{-\frac{1}{2}}\,dz$} on \HS,
the remaining distributional part of the Weyl 
curvature tensor can now be written as composed 
of the null 2--forms $\by \cart dx,\,\bz \cart dx$ as
\be
  r_{ij\,kl} = 2 \left( \by_{[i}\,n_{j]}\,\by_{[k}\,n_{l]} 
                      - \bz_{[i}\,n_{j]}\,\bz_{[k}\,n_{l]} \right) \delta(x),
\ee
clearly exhibiting its Petrov type N algebraic structure
(\ie $n^i\,r_{ij\,kl} = 0$)
--- in conformity with the Penrose type III of null junction. 
Note also, that on \HS~this expression is well--behaved (as a distribution)
near, and even `on' the string \ST,
not showing any evidence of it.
This again contrasts with the corresponding situation
of the AS-- and DH--metric, where the 
coefficient of the Weyl curvature 
distribution diverges on the null particle.

Also note that the distributional Weyl tensor is
not anymore invariant under the full $SO_3$ symmetry of the \SII~---
only by the axial rotations induced by $\vY := \prtl{Y}$.
In fact, the non--axial Killing vectors
\bea
\vK^+ &:=& +\cos(Y)\,(\vZ - 2\,\rho\sigma\tau\,\lambda\,\vW)/(1+2|X|) + \sin(Y)\,\vY \\
\vK^- &:=& -\sin(Y)\,(\vZ - 2\,\rho\sigma\tau\,\lambda\,\vW)/(1+2|X|) + \cos(Y)\,\vY,
\eea
(where $\vW := \prtl{W}$ and $\lambda := \sqrt{(1-\tau z)/(1+\tau z)}$)
are {\em discontinuous} through \HS~and 
so cannot be properly defined on it.
Anyways, the mere presence of the string \ST,
with its emphasis on $\vY$,
would already break down
the spherical symmetry to an axial one.

\section{The Continuous Potential}

Here we try to find a continuous electric potential $A_e$ for the Maxwell field
\be
F_e = \ofourp \, p\,(1+|x|)^{-2} dx \cart dw
\ee
with electric charge $p$,
which appears in the Maxwell stress tensor 
\be
M_{ij} := 4\,\pi\,\Big( g^{kl}\,F_{ik}F_{jl} - \fourth\,F^{kl}F_{kl}\,g_{ij} \Big)
\ee
in the \rhs of the Einstein equations.
We use the convention that makes the charge--integral $\oint \! \star F$ 
integer--valued when $p$ is.\footnote[3]
{therefore the ubiquituous factor $4\,\pi$ appears in the numerator instead of the denominator of $M$}
The magnetic potential $A_m$ will be dealt with separately at the end.

Just adapting the standard potential $A_e = -\ofourp \,p\,(1+|x|)^{-1}\,dw$
would not be a good choice, because due to the side--dependency of
the corresponding Maxwell field, 
$F_e = \ofourp \,\sigma p\,(1+|x|)^{-2} dx \cart dw$,
there would also be a charged shell, supported by \HS.
Also, due to the factor $dw$, it would not be continuous through \HS.
Therefore this potential is not what we are looking for.
\smallskip

\nid
A better choice is to start from the potential in the form
\be
A_e := -\ofourp \frac{p\,x}{1+|x|}\,dw,
\ee
which due to the factor $x$ is evidently continuous through \HS.
This particular form of the potential is well--known from the \MP solution, 
which is closely related to the extremal \RN solution
(\cmp \eg Hartle and Hawking \cite{HaH72}).

When expressed in the extended coordinates, this results in
\be
A_e = -\ofourp \frac{p\,X}{1+|X|} \bigg(dW - 2\,\rho\,\Lambda \, dX  
+ 2\,\frac{\rho\sigma\tau}{(1+\tau z)^2}\Big((1+\tau z)+2|X|\Big) dz \bigg), 
\ee
\bes
\where dz = \frac{dZ}{1+2\,|X|} + 2\sigma\tau\,\frac{(1-\tau z)\,dX}{(1+2\,|X|)^2} \mand z = \frac{Z+2\tau\,|X|}{1+2\,|X|}.
\ees
Here and in the following, for the sake of notational simplicity, 
we omit the indices $\rho,\,\sigma,\,\tau$ 
in the corresponding coordinate charts.
The expression for the corresponding Maxwell field is slightly simpler,
\be
F_e = \ofourp \frac{p}{(1+|X|)^2} \, dX \wedge \bigg(dW 
+ 2\,\frac{\rho\sigma\tau}{(1+\tau z)^2}\frac{(1+\tau z)+2|X|}{1+2\,|X|}\, dZ \bigg).
\ee
Note that now there is a jump in the $dZ$--term --- this will be discussed later
in the context of the \Li conditions for weak solutions.
\smallskip

\nid
The magnetic potential $A_m$ can be dealt with similarly. 
Starting first with 
\be
A_m = -\frac{q}{4 \pi}\, z\,dy,
\ee
we note, that this potential is not continuous at the poles $z=\pm1$
and so is not viable.
However, the monopole potentials
\be
A_m^\tau = \frac{q}{4 \pi}\, \tau(1-\tau z)\,dy,
\ee
are continuous either on the upper cap ($\tau=+1$) 
or on the lower cap ($\tau=-1$),
even after the replacement $z=z(Z,X)$.
They can even be made globally continuous by using the change of gauge
$A \to A + d\lambda$, $\lambda = \frac{q\, y}{2\pi}$, corresponding to the
$U(1)$--gauge transition function $S = \exp(i q\, y / 2\pi)$,
which requires the magnetic charge $q$ to be an integer, $q \in \bf N$.

The corresponding Maxwell field is then given by
\be
F_m = \frac{q}{4 \pi}\,\frac{(1+2\,|X|)\,dZ + 2\sigma\tau (1-\tau z)\,dX}{(1+2\,|X|)^2} \wedge dY,
\ee
this time showing a jump in the $dX$--term (also to be discussed later).
\smallskip

These two potentials can for our extension even merged into one
\be
A := A_e + A_m, \where q \in {\bf N} \mand p^2+q^2=m^2.
\ee

Note that up to the charge factors $p,q$ the Maxwell fields are 
exactly the Hodge--duals of each other: $q\,F_m = p \dual\!\! F_e$.
A duality rotation,
\mbox{$F \to \cos \alpha \,F + \sin \alpha \,\dual\! F$} 
with an arbitrary constant $\alpha$,
leaving the Maxwell stress tensor invariant,
is however in general not possible,
as the magnetic charge quantization would be spoiled.

Being derived directly from their standard forms,
these Maxwell fields not only satisfy 
$dF=0$ but also $\delta F = 0$ 
(where the operator $\delta := \star^{-1} d\, \star$ 
is the codifferential of $d$, depending on the metric),
and so do not involve any local currents as sources.
This is evident for $x \neq 0$.
That also no currents supported by \HS~are involved, 
will be shown next --- the Maxwell equations
thus continuing to hold in a weak sense.

\section{The Global Extension}

Up to now, we dealt only on how to obtain 
charts and transition functions describing the passage 
through a particular horizon $\Sigma_{\rho}$ (labelled by $\rho$) 
of a certain nonanalytic
extension of the exterior \RN metric and its gauge potential.
Here we want to show, how to cover completely one exterior
region, so as to reach the corresponding 
other horizon $\Sigma_{-\rho}$.
As this extension also covers $\Sigma_{-\rho}$,
the process of extension can then be continued
indefinitely both in the `future' as in the `past',
to finally arrive at the global extension.

Let us first introduce four more kinds of charts (called `standard' charts
in the following), which will be used to connect 
$\Sigma_{\rho}$ with $\Sigma_{-\rho}$.
We assume the domain of definition of the charts for the
original EF--coordinates to be given by $x > 1$, $u$ (\rsp $v$) unrestricted,
$-1 < z < +1$, and $y$ periodic in $2\pi$.
These two types of `positive' (\wrp to $x$) standard charts 
(let's just call them `$u$--chart' \rsp `$v$--chart')
completely overlap (but without covering any part of some horizon)
and are related 
by the transition functions of \eref{Tinout}.
These standard charts in turn overlap with the extended charts
with transition functions given by \eref{EF_to_X}.
Similarly, we  define two `negative' standard charts, 
now restricting $x$ by $x < -1$, 
and relate to the corresponding extended charts.
\smallskip

Let us now patch together a {\em basic building block}
(in the following, just `BBB'),
from which the complete extension can easily built up.
It essentially consists of the appropriately laid
out system of 12 charts (= 2 sets of extended charts + 1 set of standard charts)
for two exterior regions joined at a common horizon.

Start from the standard $v$--chart for an ingoing exterior RN--metric
($\rho=-1$) and consider the patches with the extended charts
$C_{-+\pm}$. Before extending this set of
charts through the future horizon, let us first extend it
into the past, by considering the corresponding
u--chart and its extended partners ($\rho=+1$). 
Now we have covered with this system of six charts
one complete exterior RN--region, including both horizons,
except for the string \ST.
As next we extend this exterior region, having $X \geq 0$, 
through the future horizon to another such exterior region,
but having $X \leq 0$, as follows:
just patch the $C_{-+\pm}$--charts
with the corresponding $C_{--\pm}$--charts
on their overlaps containing the future horizon.
Using the patches to the corresponding standard charts for $x < -1$,
they can patched to extended charts of 
type $C_{+-\pm}$.
Now our BBB is completed. 
Note that it contains three different horizons:
one joining the two exterior regions ($\rho=-1$), 
and two which still can be further extended ($\rho=+1$).
That is, we can extend ad infinitum such a BBB through their
unmatched horizons by similar BBBs.
As we did not modify topologically or metrically
the exterior RN--regions, we can be sure that
the manifold so constructed is complete 
as an extension.
This set of 12 charts can even be made
part of an maximal smooth atlas.
\medskip

\begin{figure}[h]

\setlength{\unitlength}{0.5cm}

\begin{picture}(15,16.5)(-5,-1)

\thinlines
\put( 0,10){\line( 1, 1){5}}
\put( 0,10){\line( 1,-1){5}}

\put(15, 5){\line(-1, 1){5}}
\put(15, 5){\line(-1,-1){5}}

\dashline[50]{0.4}(10,10)( 5,15)
\dashline[50]{0.4}(10,10)( 5, 5)
\dashline[50]{0.4}( 5, 5)(10,-0)

\put(10, 0){\circle*{0.25}}
\put( 5, 5){\circle*{0.25}}
\put(15, 5){\circle*{0.25}}
\put( 0,10){\circle*{0.25}}
\put(10,10){\circle*{0.25}}
\put( 5,15){\circle*{0.25}}

\qbezier[50](5,5)( 7,10)(5,15)
\qbezier[50](5,5)( 3,10)(5,15)
\qbezier[50](5,5)(10,10)(5,15)
\qbezier[50](5,5)( 0,10)(5,15)

\qbezier[50](10,10)(12,5)(10,0)
\qbezier[50](10,10)(08,5)(10,0)
\qbezier[50](10,10)(15,5)(10,0)
\qbezier[50](10,10)(05,5)(10,0)

\put(4,10){$X \leq 0$}
\put(9, 5){$X \geq 0$}

\end{picture}

\caption{\label{basic} A Basic Building Block}

\end{figure}
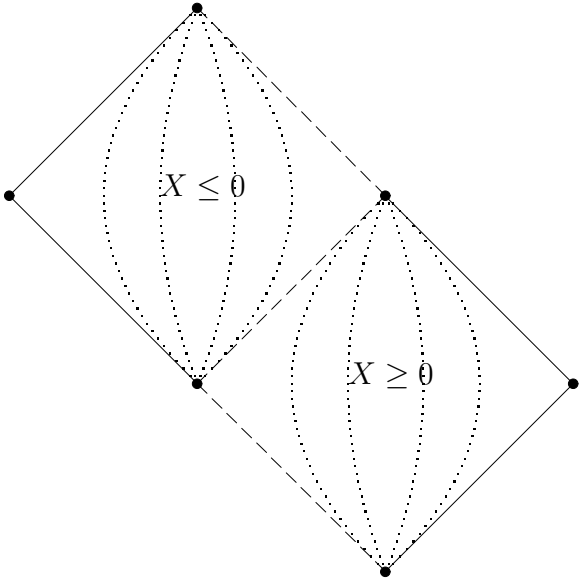

As any part of our extension not containing a horizon
is still isometric with the exterior \RN metric,
we can exhibit the BBB and the resulting manifold 
as in \fref{basic} and \fref{context} by means of 
Carter--Penrose diagrams, the broken lines
now denoting a \Li surface of nonsmoothness.
For comparison, we also display in \fref{anaext} the CP diagram of the 
standard (\ie analytic) extension.

\newsavebox{\Tclone}
\savebox{\Tclone}
{
\setlength{\unitlength}{0.25cm}

\begin{picture}(10,20)

\thinlines
\put( 0,10){\line( 1, 1){5}}
\put( 0,10){\line( 1,-1){5}}

\put(15, 5){\line(-1, 1){5}}
\put(15, 5){\line(-1,-1){5}}

\dashline[50]{0.4}(10,10)( 5,15)
\dashline[50]{0.4}(10,10)( 5, 5)
\dashline[50]{0.4}( 5, 5)(10,-0)

\put(10, 0){\circle*{0.4}}
\put( 5, 5){\circle*{0.4}}
\put(15, 5){\circle*{0.4}}
\put( 0,10){\circle*{0.4}}
\put(10,10){\circle*{0.4}}
\put( 5,15){\circle*{0.4}}

\end{picture}
}

\begin{figure}[h]

\setlength{\unitlength}{0.25cm}
\begin{minipage}[t]{8cm}

\begin{picture}(15,37)(-10,-1)

\thinlines
\multiput(0, 0)(0,10){3}{\line(1, 1){10}}
\multiput(0,10)(0,10){3}{\line(1,-1){10}}
     \put(5,35)      {\line(1,-1){5}}
\multiput(0,0)(0,0.5){70}{\circle*{0.30}}

\multiput( 0,0)(0,10){4}{\circle*{0.4}}
\multiput(10,0)(0,10){4}{\circle*{0.4}}
\multiput( 5,5)(0,10){3}{\circle*{0.4}}

\multiput(0, 0)(0,10){4}{\line(1, 1){5}}
\multiput(0,10)(0,10){3}{\line(1,-1){5}}
\put(-8,  14.5){\emph{singular}}
\put( 5.5,17.0){\emph{smooth}}
\thinlines
\put(5.0,17.5){\vector(-1,0){2.5}}
\put(-1.5,15.0){\vector(+1,0){1.5}}
\end{picture}\par
\caption{\label{anaext} Analytic Extension}
\end{minipage}
\hfill
\begin{minipage}[t]{8cm}
\setlength{\unitlength}{0.25cm}
\begin{picture}(15,37)(-2,-1)

\put(07, 0){\usebox{\Tclone}}
\put(07,10){\usebox{\Tclone}}
\put(07,20){\usebox{\Tclone}}
\thinlines
\put(13, 5){\line(-1,-1){5}}
\put(18,30){\line( 1, 1){5}}
\put(13.0,17.5){\vector(1,0){2.5}}
\put(4.5,17.0){\emph{continuous}}
\end{picture}\par
\caption{\label{context} Continuous Extension}

\end{minipage}

\end{figure}

\medskip

\section{The \Li Conditions for Weak Solutions}

The extensive and recurrent work of \Li over a period 
of almost 40 years on weak solutions of the
\MESE seems not to be very well--known --- 
in fact, nowadays almost forgotten.
Here we will necessarily remain sketchy, and refer to
\li's summarizing article \cite{Lic79} 
(\cmp also his more recent book \cite{Lic94})
for the complete theory.
\smallskip

As this concept is not yet in general use 
in the context of General Re\-la\-ti\-vi\-ty,
perhaps we should first convey the idea in a nutshell,
what is understood by a {\em weak solution}
to some system of quasi--linear second--order partial equations,
like the \MESE is. 
Assuming a continuous metric 
(which it should be on geometrical and physical grounds)
and gauge potential, their first derivatives
could jump on certain submanifolds,
which turn out to be characteristics,
\ie null hypersurfaces \HS. The second
derivatives appearing linearly in the
field equations can then still be
interpreted in the sense of distributions,
giving rise to Dirac--type of distributions
supported by \HS.
In contrast to some other approaches,
which start from manifestly distributional metrics
(\cmp \eg the survey in \cite{GHK99}),
we can stay in the well--established realm of classical,
\ie {\em linear}, distribution theory \`a la Schwartz.

Now, a weak solution is a solution to the
basic field equations, which now are interpreted
in the weak sense, \ie with generalized
derivatives potentially resulting in Dirac--like
distributions. However, as the basic field equations
are otherwise not modified, any possible distributional
term in them must be cancelled.
For the \MESE this implies the following:
a) the Maxwell field is devoid of charged shells,
b) in the Einstein equations, which on the \rhs
are quadratic in the Maxwell field 
(which is of first order in the potential)
any distribution can only enter through
the Riemann curvature on which the \lhs is based ---
in particular, no material shell should remain.
This is the case, when 
any integral of the field equations multiplied
with some test function vanishes with respect to 
the support of the test function.
This is the essence of the concept of a weak solution.
\smallskip

Note however, that in the context of the
\ME equations
some manifestly distributional
expressions of Dirac--type can still remain --- this is the case
of the Weyl component of the curvature.
For this reason, such solutions are since
Penrose also called {\em impulsive gravitational waves}
(in contrast to {\em gravitational shock waves},
where there are only jumps in the curvature).
Despite basic differences in their respective mathematical approaches,
from the point of view of physics,
they should be considered as fundamentally the same
as the weak solutions in the sense of \li.
\smallskip

Also note, that the metrics of \AaS and of \DtH
fall outside this class of solutions in the {\em strict sense}:
for one thing, being manifestly distributional,
they are not continuous, and for the other,
the (vacuum) Einstein equations do not hold
in the weak sense, as appropriate distributional
terms have to be explicitly introduced on the \rhs as sources.
Also the cone metric, as analyzed by Clarke et al.~\cite{CVW96}
in terms of Colombeau distributions,
falls into the same category.
\medskip

Now, the {\em \Li conditions}\footnote[1] 
{not to be confounded with his better--known {\em junction conditions}}
to establish the weakness of 
certain solutions of the \MESE consist of two parts:
\bit
\item [i)] of {\em technical assumptions}
(like appropriate differentiability of the local description of the 
null hypersurface \HS, 
the uniform continuity of the first derivatives
of metric and gauge potential and the choice 
of an appropriate space of test functions),
guaranteeing the mathematical existence
of the objects under discussion 
(\eg second derivatives as Dirac distributions supported by a null hypersurface)
\item [ii)] conditions referring to the {\em jumps} of
certain derivatives of the {\em metric and gauge potential},
in order that the resulting solution can be interpreted 
as a weak solution of the \MESE
\eit
As for our extension, which admits an expansion in $x$
both for $x\geq0$ and $x\leq0$ separately, 
the technical assumptions i) are easily checked to hold, 
we will deal exclusively with the jump conditions ii).
They consist in some algebraic restrictions on the jump
quantities, to prevent the appearance of distributional terms
in the weakly interpreted (\ie using generalized derivatives)
field equations.

These jump conditions can be derived and formulated
most easily for the Maxwell field.
Consider the continuous function $f(x) := |x|$.
Its derivative is essentially the (discontinuous) 
sign--function: $df = \mbox{sign}(x)\,dx$.
Defining $g(x) := \mbox{sign}(x)$, 
then evidently $\jump{g}=2$, and its (generalized)
derivative in turn is $dg = 2\,\delta(x)\,dx$.
Consider now a two--form $F$, with $dF=0$
and $d\dual F=0$ for $x\neq0$ and jump $\jump{F}$
at $x=0$, then we have
$dF = n \wedge \jump{F}\,\delta(x)$ and
$d\star F = n \wedge \jump{\star F}\,\delta(x)$.
Therefore, to avoid any currents supported by the
hypersurface defined by $x=0$,
we have to require $n \wedge \jump{F} = 0$ 
and $n \wedge \jump{\star F} = 0$.
Dualizing the last equation,
the \lhs is equivalent to $\vn \cont \jump{F}$
and we finally get as the {\em \Li jump conditions for
the Maxwell field}
\be
\label{lic_max}
n \wedge \jump{F} = 0 \mand \vn \cdot \jump{F} = 0,
\ee
where $n$ is the normal 1--form $dx$ of the 
hypersurface of discontinuity \HS,
and $\vn$ the corresponding vector.
It can be shown, that for an effective jump 
(\ie $\jump{F}\neq0$), this implies $\vn \cont n = 0$
--- in other words the null character of the hypersurface \HS.
\smallskip

Let us now verify these conditions for the
Maxwell fields of our extension;
to wit $\jump{F_e} \sim dX \cart dZ,\:\jump{F_m} \sim dX \cart dY$.
In fact, since $n \sim dX,\:\vn \sim \prtl{W}$,
they are evidently identically satisfied.
\medskip

The {\em \Li jump conditions for
the gravitational field} we have already formulated
and used without explicitly naming them so.
As they are notationally slightly more complex,
let us introduce the jump tensor of the $x$--derivative
of the metric, $h := \jump {g\,'}$,
which is a symmetric tensor defined on \HS.
Define the coefficient of the curvature distribution
\be
r_{ij\,kl} := n_{[i}\,h_{j][k}\,n_{l]}.
\ee
Then the Ricci--distribution vanishes 
(and for dimension $n>2$ also the corresponding 
Einstein distribution 
$e_{ij} := r_{ij} - \half \, g^{st}\,r_{st}\,g_{ij}$) {\em iff}
\be
g^{ik}\,r_{ij\,kl} = 0,
\ee
which is one of \li's conditions.
As it is the most relevant one, let us call it the
{\em primary} condition.
The other conditions of \Li are formulated in analogy to
the jump conditions for the Maxwell field and are
\be
r_{ij\,[kl} \,n_{m]} = 0 \mand r_{ij\,kl} \,n^l = 0.
\ee
Evidently the first one is vacuous,
as it is by construction identically satisfied.
The second is equivalent to demanding Petrov type N
of the Weyl distribution, when the Ricci distribution vanishes
(which is the content of the primary condition).
However, as Penrose has shown,
Petrov type N of the Weyl distribution is an automatic
consequence of Penrose type III of the junction,
which only requires $r_{ij} \sim n_i\,n_j$ for the 
Ricci distribution.
Therefore, in the context of null junctions,
$\vn \cont n =0$, the second of the extra conditions
turns out to be superfluous --- so let's call
these conditions {\em secondary}.

Assuming for \HS~a {\em null} normal 1--form $n = dx$
(which is guaranteed to be the case 
either when a nonvanishing weak Maxwell field is present
or when the secondary conditions hold),
we can put the primary condition into an equivalent 
simpler form as follows (\cmp Graf \cite{Gra97}).
Defining the {\em classifying vector} $\xi$\footnote[1]
{this vector appears already in the work of Dautcourt \cite{Dau64}
on material shells}
of the null junction as
\be
\xi^i := g^{ir}\,h_{rs}\,n^s - \half\,g^{rs}h_{rs}\,n^i,
\ee
the Ricci distribution can also be expressed as 
$r_{ij} = \xi_i\,n_j + \xi_j\,n_i$.
Evidently, the vanishing of the Ricci distribution $r_{ij}$
(eo ipso, Einstein distribution) is equivalent
to the vanishing of the characteristic vector $\vxi$.
Therefore, for a null junction,
the complete set of Lichnerowicz conditions for the jumps of the
first derivatives of the metric
reduces to the {\em single} jump condition
\be
\xi^i = 0,
\ee
which is even more manageable than his primary condition alone.
\smallskip

For our extended metric, and using the jump expressions \Ref{metric_jumps},
this jump condition is evidently satisfied.
\medskip

Skipping the technical prerequisites
(which however can easily shown to be satisfied),
both the metric and the potential (over its field tensor)
of our extension satisfy the complete set of
\Li jump conditions.
As a consequence, the null shell \HS~does not support
any charges, nor any material sources. Concluding,
{\em our extension can be considered to be
a solution of the \MESE in the weak sense
of \li.}
\medskip

\nid
Some additional remarks are in order.
\smallskip

\nid
For one thing, in the general non--weak case,
the Einstein distribution $e$ is easily seen 
to obey the ubiquitous {\em Lanczos identity}
$\vn \cont e = 0$, when expressed by means of 
$\xi$ as $e_{ij} = \xi_i\,n_j + \xi_j\,n_i - (\xi \cont n)\,g_{ij}$
\smallskip

\nid
However, for a weak \ME solution, the Einstein distribution
of course vanishes. Nevertheless, there remains
a similar identity at the level of the discontinuity
(expressed as jump) of the Einstein {\em tensor} $E_{ij}$,
which we want to derive now. 
For this purpose we must first introduce the 
following two operators for a quantity discontinuous
at \HS.
\be
\avrg{f} := \half\, (f_+ + f_-) \mand \jump{f} := \case12\, (f_+ - f_-).
\ee
It can easily verified that they obey the following product rules
for a product of quantities discontinuous at \HS,
\bea
\avrg{f\,g} &=& \avrg{f} \avrg{g} + \jump{f} \jump{g} \\
\jump{f\,g} &=& \jump{f} \avrg{g} + \avrg{f} \jump{g}.
\eea
The particular simple form of these product rules justify 
the uncommon factor $\half$ in the definition of the jump operator.
With $\jump{F_e} \sim dX \cart dZ$, $\avrg{F_e} \sim dX \cart dW$, 
     $\jump{F_m} \sim dX \cart dY$ and $\avrg{F_m} \sim dZ \cart dY$, 
we then get for the jump of the Maxwell stress tensor $\jump{M} \sim dX \, dZ$.
Due to the Einstein equations $E \sim M$, 
we then have also $\jump{E} \sim dX \, dZ$.
This jump is obviously annihilated by contraction with the 
normal vector $\vn \sim \prtl{W}$: 
$\vn \cont \jump{E} = 0$, and so again a Lanczos identity holds
--- this time for the {\em jump} of the Einstein tensor.
In fact, in this form it is the classical identity, 
derived long ago by Lanczos \cite{Lan22}, \cite{Lan24}.
Note that for solutions being only gravitational shock waves
(\ie \CI~metrics, with only jumps in the curvature), this Lanczos identity
can be derived more directly, and without using the Einstein equations
(\cmp \eg Hamoui and Papapetrou \cite{HaP68}).
\smallskip

\nid
Incidentally, for the extreme \RN metric,
the Weyl tensor $W_{ij\,kl}$ can be shown to behave
as $o(|x|)$ near \HS, and so it vanishes there.
This has as consequence, that even when extended,
there are no jumps of it supported by \HS.
Disregarding the distributional component,
it could therefore be considered as continuous there.
\medskip

Concluding, for our weak solution we then have the following situation
regarding the total curvature tensor near \HS:
only the Weyl component is manifestly distributional
and otherwise continuous, whereas the Einstein component
has only a discontinuity at \HS, but obeying a Lanczos identity.
Regarding the Maxwell field, we only can say that it is discontinuous at \HS,
but respecting the \Li jump conditions.

\section{Conclusions}

We have shown that a continuous extension of the exterior region
of the extremal \RN metric through any of its horizons
is possible, although with a change of topology
with the physical interpretation of closed null strings.
Moreover, this extension can be continued globally
to form an inextensible \CINF~manifold.
This extension can even be understood as a solution
of the \MESE in the weak sense of \li.
Except for horizon--supported Dirac distributions
in the Weyl tensor, no distributions of 
Dirac type (or derivations thereof) appear.
\smallskip

The question as to the stability of such a solution should be reconsidered, 
especially in view of the negative result of Simpson and Penrose \cite{SiP73}
concerning electromagnetic perturbations of a \RN metric with $e^2 < m^2$.
However, some early investigations of Hajicek \cite{Haj81} seem to indicate, 
that extremal solutions have better chances to be stable.
In any case, such `stringy' topologies, 
together with nonanalytic behaviour of metric
and gauge potential,  should be taken into account in the
corresponding numerical approximation schemes ---
the `locus' of the string now introducing
some extra degrees of freedom.
\medskip

The methods developed here and applied to extension of the extremal 
\RN metric can be extended in several directions:
\bit
\item[i)]   classify the globally inequivalent continuous extensions
            of the eRN metric
\item[ii)]  deal with `nonstatic' joins and apply them to the general 
            (\ie non--extremal, non--naked) \RN solution
\item[iii)] apply them to the extremal Kerr (\rsp Kerr--Newman) solution,
            its classification, as well as the general case.
\eit
Research along these lines is being pursued.
However, it already appears that only the extensions of the extremal metrics
admit an interpretation as weak solutions in the sense of \li.
\smallskip

Of course, the extremal case, with mass $m = |e|$ of the order 
of the Planck mass,
is already within a regime where quantum gravity effects are believed 
to be essential.
Therefore the direct physical relevance of the present work 
may be questioned. However in the context of String Theory,
extremality seems to be a relatively common feature.
After all, we showed that already in the classical context of a
continuous extension of the extremal \RN metric,
a Null String seems to be more natural ---
at least, better behaved ---
than a Null Particle.

\section*{Appendix}

Here a somewhat more detailed derivation of the 
distributional part of the stress--energy tensor
for the `naive' metric will be given.
This derivation can straightforwardly be extended also to
more complicated metrics depending on nonsmooth expressions like $|x|$, 
as for the continuously extended extremal \RN metric.

As both the mixed Riemann and covariant Ricci tensors have the 
symbolic form
\( R \sim \Gamma^2 \oplus \partial \Gamma \),
with the the affinities
\( \Gamma \sim g^{-1} \otimes \partial g \),
for a metric depending on the nonsmooth expression \( |x| \),
a delta distribution can only be expected from the second partial 
derivatives of the metric \( g \). 
An explicit expression for the totally covariant Riemann tensor
can be found in some older textbooks like  Synge \cite{Syn60},
giving for the terms with second derivatives\footnote[3]
{throughout this appendix we will use the definitions and sign conventions of Synge ---
the resulting expression for the stress--energy tensor $E$ will then agree 
effectively with the commonly used one}
\bes
R_{ij\,kl} := \half \, \Big(g_{il,jk} + g_{jk,il} - g_{ik,jl} - g_{jl,ik}  \Big) + \cdots \ .
\ees

The relevant $x$--dependent metric components for the naive metric evidently are 
\( f(x) := g_{w w} = -x^2/(1+|x|)^2 \) 
and the coefficient of \( d\Omega^2 \), given by \( g(x) := (1+|x|)^2 \).
Only the latter has a second distributional derivative containing a delta 
distribution, since by formally deriving the absolute--value function $|x|$
and using the distributional identity $|x| = x\,\mbox{signum}(x)$, we get
\beas
f'\,(x) &=& 2 \, x / (1+|x|)^3, \qquad \qquad \qquad     g'\,(x) = 2\,(x + \mbox{signum}(x)), \\
f'' (x) &=& 2 \, (1 - 2\,|x|) / (1+|x|)^4, \qquad  \ \,  g'' (x) = 2 + 4\,\delta(x).
\eeas
This results in the two essentially nontrivial distributional curvature components
\beas
R_{xy\,xy} &=& -4\,\delta(x) \, (1-z^2) \\
R_{xz\,xz} &=& -4\,\delta(x) \, (1-z^2)^{-1} \ .
\eeas
For the only nontrivial distributional component of the  Ricci tensor
$R_{jk} := g^{il} \, R_{ij\,kl}$ we then get 
\bes
R_{xx} = 8 \, \delta(x).
\ees
As the corresponding scalar part $R$ of the Ricci tensor vanishes, 
we get for the Einstein tensor $G_{ij} := R_{ij} - \half \, R\, g_{ij}$,
exactly the same expression $G_{xx} = 8 \, \delta(x)$ for
its distributional part.
Using the field equations in their form $G_{ij} = -8\,\pi\,E_{ij}$,
we finally get 
\bes
E_{xx} = -\oonep \, \delta(x),
\ees
which describes null dust matter with negative energy density. 
Our equation \eref{dis} in the main text
is just another way to express $E$.
\medskip

For more complicated metrics depending on $|x|$, 
like the continuously extended extreme \RN metric
of section \ref{ceern_metric},
exactly the same procedure can be applied to
calculate the distributional part of
the stress--energy tensor.
A major simplification can be achieved by first disregarding 
in the metric all the $o(x^n)$--terms with $n > 1$,
as they are completely irrelevant for the result.

Also this method can be easily adapted to metrics
depending on $x$ through the more commonly used $f(x) := x\,H(x)$--terms,
where $H(x)$ is the Heaviside step--function.

\section*{Acknowledgments}

I wish to thank Peter Aichelburg and Herbert Balasin
for their useful comments and suggestions.


\newpage

\begin{thebibliography}{99}

\bibitem{Lan22}
Lanczos K 1922 {Bemerkung zur de~Sitterschen Welt}.
{\it Phys.~Zeitschr.}, {\bf XXIII} 539-543

\bibitem{Lan24}
Lanczos K 1924 {Fl\"achenhafte Verteilung der Materie in der Einsteinschen 
Gravitationstheorie}. {\it Ann.~d.~Phys.}, {\bf 74} 519-540

\bibitem{Syn60}
Synge J L 1960 {\it Relativity: the General Theory.}
(Amsterdam: North--Holland Publ.)

\bibitem{Whe62}
Wheeler J A 1962 {\it Geometrodynamics.}
(New York: Academic Press)

\bibitem{Dau64}
Dautcourt G 1964 
{\"Uber Fl\"achenbelegungen in der allgemeinen
  Rela\-tivi\-t\"ats\-theorie}.
{\em Math.~Nachr.}, {\bf 27} 277-288

\bibitem{Car66}
Carter B 1966 The complete analytic extension of the \RN metric 
in the special case $e^2=m^2$.
{\it Phys.~Lett.}, {\bf 21(4)} 423-424

\bibitem{HaP68}
Papapetrou A and Hamoui A 1968 Couches simples de mati\`ere en relativit\'e g\'en\'erale.
{\it Ann.~Inst.~Henri Poincar\'e}, {\bf IX(2)} 179-211

\bibitem{AiS71}
Aichelburg P C and Sexl R 1971 On the gravitational field of a massless particle.
{\it Gen.~Rel.~Grav.}, {\bf 2(4)} 303-312

\bibitem{Pen72}
Penrose P 1972 The geometry of impulsive gravitational~waves.
{\it General~Relativity. Papers in Honour of J.~L.~Synge}, pages 101-115. 
ed Raife\-ar\-taigh L O (Oxford: Clarendon Press)

\bibitem{HaH72}
Hartle J B and Hawking S W 1972 Solutions of the Einstein--Maxwell Equations
with Many Black Holes.
{\it Comm.~math.~Phys.}, {\bf 26} 87-101

\bibitem{SiP73} 
Simpson M and Penrose R 1973 Internal Instability in a \RN Black Hole.
{\it Int.~J.~of Theor. Phys.}, {\bf 7(3)} 183-197

\bibitem{Eat78}
D'Eath P D 1978 High-speed black-hole encounters and gravitational radiation.
{\it Phys.~Rev.~D}, {\bf 18} 990-1019

\bibitem{Lic79}
Lichnerowicz A 1979 Relativity and Mathematical Physics.
{\it Relativity, Quanta, and Cosmology},
volume~II, ed de Finis F, chapter~12, pg.~403-471. 
(New York: Johnson Reprint Co.)

\bibitem{Haj81}
Hajicek P 1981 Quantum Wormholes (I). Choice of the classical solution.
{\it Nucl.~Phys.~B}, {\bf 185} 254-268

\bibitem{DrH85}
Dray T and 't~Hooft G 1985 The gravitational shock~wave of a massless particle.
{\it Nucl.~Phys.~B}, {\bf 253} 173-188

\bibitem{ClD87}
Clarke C J S and Dray T 1987 Junction conditions for null hypersurfaces.
\CQG {\bf 4}, 265-275

\bibitem{GeT87}
Geroch R and Traschen J 1987
Strings and other distributional sources in general~relativity.
{\it Phys.~Rev.~D} {\bf 36(4)}, 1017-1031

\bibitem{Gri91}
Griffiths J B 1991 {\it Colliding Plane Waves in General~Relativity}.
(Oxford: Clarendon Press)

\bibitem{BaN93} 
Balasin H and Nachbagauer H 1993
Distributional Energy--Momentum Tensor in the Kerr--Newman Space--Time Family.
\CQG {\bf 11}, 1453

\bibitem{Lic94}
Lichnerowicz A 1994 {\it Magnetohydrodynamics: 
Waves and Shock Waves in Curved Space--Time.}
(Amsterdam: Kluwer Academic Publ.)

\bibitem{CVW96} 
Clarke C J S, Vickers J A and Wilson J P 1996 
Generalised Functions and Distributional Curvature of Cosmic Strings.
\CQG {\bf 13}, 2485-2498

\bibitem{AiB96} 
Aichelburg P C and Balasin H 1996 
Generalized Symmetries of Impulsive Gravitational Waves.
\CQG {\bf 14}, A31-A42

\bibitem{Bal97} 
Balasin H 1997
Distributional energy momentum tensor of the extended Kerr geometry.
\CQG {\bf 14}, 3353-3362

\bibitem{Gra97} 
Graf W 1997
A Classification for Null $C_r$-Junctions ($r \geq 0$).
{\it unpublished}

\bibitem{AiB98} 
Aichelburg P C and Balasin H 1998 
ADM and Bondi four--momenta for the ultrarelativistic \Sc black hole.
\CQG {\bf 15}, 3841-3844

\bibitem{PoV98} 
Podolsk\'y J and Vesel\'y K 1998
Continuous coordinates for all impulsive pp-waves.
{\it Phys.~Lett.}, {\bf A241} 145-147

\bibitem{Car99}
Carlip S 1999
Entropy from Conformal Field Theory at Killing Horizons.
{\it Preprint} gr--qc/9906126

\bibitem{GHK99} 
Grosse M, H\"ormann G, Kunzinger M and Oberguggenberger M (eds.) 1999
{\it Nonlinear Theory of Generalized Functions}.
(London: Chapman \& Hall)

\bibitem{AiB00} 
Aichelburg P C and Balasin H 2000
Generalized asymptotic structure of the ultrarelativistic \Sc black hole.
\CQG {\bf 17}, 3645-3662

\bibitem{Bal00} 
Balasin H 2000
Generalized Kerr--Schild metrics and the gravitational field 
of a massless particle on the horizon.
\CQG {\bf 17}, 1913-1920

\bibitem{Ste00} 
Steinbauer R 2000
Distributional Methods in General Relativity.
{\it PhD thesis, University of Vienna}

\bibitem{Ash00}
Ashtekar A 2000 
Interface of General Relativity, Quantum Physics 
and Statistical Mechanics: Some Recent Developments.
{\it Ann. Phys. (Leipzig)}, {\bf 9(3-5)} 178-198

\bibitem{ABF00}
Ashtekar A, Beetle C and Fairhurst S 2000
Mechanics of isolated horizons.
\CQG {\bf 17}, 253-298

\bibitem{ABL01}
Ashtekar A, Beetle C and Lewandowski J 2001
Mechanics of Rotating Isolated Horizons.
{\it Preprint} gr--qc/0103026 v2

\bibitem{JaM01}
Jassal H K and Mukherjee A 2001
Classical string propagation in gravitational fields.
{\it Preprint} hep--th/0106164

\bibitem{SjV01}
Sj\"odin K R P and Vickers J A 2001
The thin string limit of Cosmic Strings coupled to gravity.
{\it Preprint} hgr--qc/0108009


\end{thebibliography}
\end{document}